**From private to public governance: The case for reconfiguring energy systems as a commons**

*Chris Giotitsas*

Ragnar Nurkse Department of Innovation and Governance

Tallinn University of Technology (TalTech)

*Pedro H. J. Nardelli*

School of Energy Systems

Lappeenranta-Lahti University of Technology

*Vasilis Kostakis*

Ragnar Nurkse Department of Innovation and Governance

Tallinn University of Technology (TalTech)

Berkman Klein Center for Internet & Society

Harvard University

*Arun Narayanan*

School of Energy Systems

Lappeenranta-Lahti University of Technology

## Abstract

The discussions around the unsustainability of the dominant socio-economic structures have yet to produce solutions to address the escalating problems we face as a species. Such discussions, this paper argues, are hindered by the limited scope of the proposed solutions within a business-as-usual context as well as by the underlying technological rationale upon which these solutions are developed. In this paper, we conceptualize a radical sustainable alternative to the energy conundrum based on an emerging mode of production and a commons-based political economy. We propose a commons-oriented Energy Internet as a potential system for energy production and consumption, which may be better suited to tackle the current issues society faces. We conclude by referring to some of the challenges that the implementation of such a proposal would entail.

## 1. Introduction

The last decade has been pivotal for elevating in the global conscience the need for radical changes in the “way we live”. There have been voices that demanded such changes before, but only recently has it reached a critical mass, galvanized by a climate crisis that may threaten our very existence on the planet [1] and the recent outbreak of COVID-19 [2]. Amidst these discussions for change, the issue of rethinking our energy systems–traditionally centralized and fossil-fuel-based to support mass consumption–has been prominent and contentious, with no clear solutions on the horizon while critiques of the incumbent systems intensify.

This paper does not contribute to the critique as the latter is extensive and robust. Instead it tries to conceptualize radical sustainable alternatives to the energy conundrum based on an emerging mode of production and a commons-based political economy. Too often energy research is removed from the social sciences and thus becomes limited within the technical boundaries of the field [3]. Sovacool et al. [4] point out how many published studies fail to address novel and interesting research questions as well as to apply rigorous research methods. We wish to build on the former point and suggest that for truly impactful energy research now, it is imperative to examine solutions that challenge the very foundation of modern organisational and production systems. We explore a type of institutional innovation, which is “important in the light of climate change as a commons problem” [5]. The aim is to contribute to a rhetoric and a sociotechnical imaginary of energy transitions [6].

The model proposed in this paper is a theoretical exercise merging current research in distributed energy production and microgrids with action research on the potential of technology development within a commons framework. It constitutes an attempt to provoke debate and invite more scholars (both in engineering and social sciences) and apply those rigorous research methods to explore this proposed, or any other radical, alternative to the energy conundrum. Hence, while the framework proposed here is comprehensive in technical terms, it is galvanized by normative research on radical social transformation.

The paper is organized as follows: Section 2 discusses how current proposed solutions are hindered by addressing the problem in a business-as-usual framework from a political economy perspective. Section 3 elaborates on an alternative, commons-based framework within which new solutions may thrive. Section 4 presents the commons-oriented Energy Internet proposal in practical terms. Section 5 concludes by pointing to the political challenges such a proposal would entail.

## 2. Technology in capitalism

The discussions around the unsustainability of current socio-economic structures have yet to produce solutions that would address the escalating problems we face as a species. Such debates appear to be hindered by two overarching issues. First, the scope of the proposed solutions is limited within a business-as-usual context with regards to a political economy driven by compound economic growth. And, consequently, the underlying technological rationale upon which these solutions are developed is similarly restricted. Allow us to elaborate.

Developed and developing capitalist economies are faced with unprecedented threats that stem from the ways they organize, produce and consume. Capitalism is a hyper-productive system based on extractivism, compound economic growth and profit-maximisation-oriented structures of power [1]. Nature is considered a commodity to which we are entitled to rather than us being a part of a complex and delicate ecosystem [7,8,9]. All interventions and extractions, in this regard, are mere externalities in the equations that determine the next logical step to economic growth. The problem with this system is that the planet's resources are finite and the disregard for its ecosystems has brewed significant and, perhaps, irreparable consequences.

It is this unfettered commodification, accumulation, compound growth and consumption that is the basis of the capitalist system that renders the solutions offered inadequate. One cannot do away with these aspects without changing the system itself [10, 11]. Applied to the energy systems conundrum, it is easy to see how these systemic imperatives prove to be massive roadblocks for meaningful progress. The dependence on fossil fuels has not decreased substantially for a number of reasons. We argue that it is not a matter of technological infrastructure or even available resources. But it is due to economic and political reasons that the transition to other sources of energy has been slow, with massive energy conglomerates protecting their interests by preventing policy which would benefit alternatives as well as the potential incompatibility of renewable energy and profitable investments [12, 13, 14]. Though undoubtedly spatial, resource and energy demand as well as consumption considerations also come into play [15, 16]. At least, that is, considering the current socio-economic system and its demands.

However, even if these issues are overcome, the severity of the current predicament makes it palpable that a whole host of new problems need to be tackled with the so-called renewable sources of energy. First, the technologies themselves are reliant on rare and non-renewable materials, typically mined under unsustainable and inhumane conditions in marginalized regions in Africa and Asia [17]. Large scale implementation of renewable energy production would merely displace environmental damage elsewhere on the planet. Furthermore, such installations may cause significant adverse effects on both the environment as well as local communities [18, 19, 20].

These business-as-usual approaches to tackle the issue highlight the social imperatives that limit the capacity to conceptualize innovative technological solutions. Broadly speaking, there are two wider philosophical viewpoints on technology development. On the one hand, technological determinism maintains that society is deeply affected by technological change, while technology itself follows its own inescapable path [21]. On the other, social determinism views technology as the outcome of complex social interactions both at the micro- and macro-level.

Most technology theorists align to the latter side with varying degrees of vehemence following a plethora of empirical studies that track the social origins for the shape of multiple technological artefacts (for an extensive overview, see [22]). Yet when it comes to large technological systems, the truth lies somewhere in the middle. At its core, technology is indeed the product of human creativity; it does not have a mind of its own. However, once technology reaches a certain level of complexity, which comes with massive material infrastructure and socioeconomic attachments, it gains significant “momentum” that can greatly affect how society evolves to adapt around that infrastructure [23]. Energy systems are a prime example and have been examined as such extensively [24, 25].

So, it becomes obvious why one would believe that technology is one of the most important factors for social change. And why the deterministic view still holds significant power in society; it corresponds to peoples’ experience with technology [21]. However, this acceptance of technology simply progressing according to scientific discovery does not allow for meaningful critique of current options or the development of radical alternatives.

On a grander scale, technology follows a similar trajectory to that of the incumbent socio-political system. The system imbues the underlying technical rationale with its dominant values [26, 27]. Thus, for more than a century, the dominant technological trajectory has been the capitalist one with its values of profit maximization, economic efficiency and power dominion of certain social groups [26]. Even self-proclaimed communist regimes imported technological and managerial methods that mirrored the capitalist ones [28]. In that sense, current proposals to tackle the energy problem are not only limited in economic and political scope, but also on a technical level. Marcuse criticized technical rationality in capitalism for its many irrationalities (like social and environmental degradation) and proposed that a new historical subject (something akin to a catalyst or an agent) needs to be established for transformation to be achieved [29]. In the context of this paper, this would be alternative technological trajectories imbued with a different set of values.

It is difficult to envision how these technologies can be developed outside capitalist frameworks because their role would not be generative but remediative; not a profit but a public/environmental good. Yet, such technological trajectories have been emerging across the planet in multiple productive fields like software, agriculture and even energy [27].

## 3. A new paradigm?

With the advent of information and communication technologies (ICTs), a novel mode of production has been emerging. Individuals across the globe have been empowered to collaborate and manufacture locally. Initiatives in multiple productive sectors have appeared, starting from software and content production, with free and open-source software and Wikipedia, and expanding to fields of physical production, like agriculture, medical applications and even space research [30, 27, 31]. These initiatives can be considered the seedlings for the mode of production known as commons-based peer production. Commons-based peer production describes the Internet-enabled free engagement and cooperation of the people, who coalesce to create shared value according to community-defined governance mechanisms [30, 32].

This mode of production differs from the capitalist one in several key ways. Profit and accumulation are not the main drivers for production [30, 32]. The economic sustainability of those involved is important, but the imperative of economic maximization is set aside. Collaboration is the goal rather than a mere tool and the output is a commons to be shared according to rules and norms defined by the producers and based on needs and use. This means that in the case of knowledge production, the product itself is the commons: a piece of software, music, video or just informative data. For manufacturing, the design and all relevant information of creating and using the product are made available. Moreover, certain communal spaces may be organized to offer the tools for the actual creation. Examples abound, from agricultural machinery and small-scale manufacturing to delicate medical applications and energy production equipment [27, 31].

Thus, the values that are embedded in such a mode of production are often suppressed in the dominant mode and are as diverse as the initiatives themselves. However, some of these values can be traced across the board like openness of knowledge and processes, autonomy in terms of organization and resource management and sustainability both economic and ecological. For these initiatives, and the radical technology they envision, to move from the fringes of production in society and into the center of activity, a boost would be necessary. The commons literature often presents them as the third way between private and public sectors [6]. However, for a transition into the third option to be achieved, not only would the state have to exhibit a non-hostile disposition, but it would have to assist it as well.

Commons scholars [32, 33] have proposed a new form of symbiosis between state and civil society, based on the principles and practices of commons-based peer production, as an extension of the global-western Welfare State. The idea of the Welfare State is tuned with capitalist production focusing on redistributing wealth to alleviate externalities, such as income inequality and environmental degradation [33, 34]. The proposal is to shift from redistribution to pre-distribution, harnessing the productive capacities around the commons. The externalities could be internalized by embedding productivity within social and ecological limits, defined by commoning [33].

Although it appears that the capitalist system hindering the energy sector had major contributions to commons-based communication networks, there are important differences both in the nature of data and energy as a resource that has also affected their historical trajectories. Data is a malleable resource, and hence, the communications sector is based on capitalist structures taking advantage of inherent peer-to-peer (P2P) sharing potential; indeed, P2P sharing is the foundation of the modern Internet. On the other hand, electricity has unique properties that lead to many more technical boundaries, such as frequency control and voltage regulation. As a result, in the electricity sector, P2P sharing has always been a fringe element with centralized architectures being used to ensure that electricity is delivered without disruptions, often as a public good. Today, with the transition to open competitive electricity markets, that public good has been converted to a private good within a capitalist framework that hinders the disruption needed for a quick transition to high renewable energy penetration.

The next section presents a proposal for an energy transition specific to energy production. A transition that would require radical changes in the political and economic principles that govern the sector but for which the technological infrastructure is available, and its feasibility appears realistic.

## 4. Energy as a commons

### *4.1 A changing energy sector*

Electricity as an energy sector dates back more than a century ago and there are multiple scientific accounts on the evolution of the technology and the relevant infrastructure [35]. In its early stages, the electrified energy system was based on local generation for local usage. Technological developments and the nature of electric phenomena allowed for low-cost long-distance transmissions and opened the path for a big boom in electricity networks. The role of the state was pivotal with massive investment programs targeting “electrification” after World War II [36]. As technology developed further, still with strong state support, the electricity network became increasingly interconnected and designed to work in a centralized fashion [37].

Since the last three decades, the rule of thumb has been that electricity is governed by liberalized markets that target accumulation and profits for private actors via the supply-demand mechanism. In this context, the operation of the grid is unbundled, and for-profit entities have emerged in all sectors: large-scale generators, retailers, and regulated monopolies like distribution and transmission system operators and, more recently, ancillary service providers and aggregators [38]. Such market liberalization and deregulation had serious pitfalls, as exemplified by California’s energy crisis in 2001-2002 [39].

Today, we are witnessing a return to decentralized energy systems, where large-scale generation transmitted unidirectionally over long distances to end-consumers is supplemented, and sometimes even replaced, by local solar and wind generation, together with thermal and battery storage at low voltage distribution networks (i.e., closer to the end-point edge of the traditional centralized network). New solutions to support the integration of those distributed applications and coordinate their connections to the main grid are being researched and developed.

Two key technologies are worth mentioning here: software-defined energy networks (SDEN) [40] and packetized energy management (PEM) [41]. Enabled by the ICT, these technologies enable a computationally light but operationally efficient rule-based energy resource allocation. To demonstrate their functions and the possibilities, we draw a parallel with communications networks. Before the Internet, communication networks (e.g., telephone lines) would work with circuit-switching technologies, which basically reserved the cable for only one transmission over time. This means that the full cable capacity ("the circuit") was reserved, even when nothing was transmitted. When we called someone, the multiple "silent" moments during the conversation led to under usage of infrastructure, and increased delays/blockages for other persons willing to use the telephone. The same was true with circuit-switching-based data transmissions beyond voice. With this technology, the Internet as we know it today would probably not exist.

The radical change occurred with a technology called packet switching, which revolutionized communication networks by its novel use of multiplexing. In simple terms, big data packets could be divided into smaller blocks or packets that would not need to be sequentially transmitted nor go through the same physical route. Now the final message could be reconstructed at the receiver side. So, a better utilization of the "physical" infrastructure was now possible including definition of priorities and quality requirements. We foresee PEM to bring the same level of change in energy systems.

### *4.2 Diverging trajectories of technological development*

Much like with the Internet, multiple paths and potentialities exist to radicalize the energy system by using an "Energy Internet". The Energy Internet refers to a large-scale cyber-physical system that uses PEM of flexible loads in microgrids, enabled by the advances in ICT [42]. While we acknowledge that only a small subset of the solutions presented so far in energy systems literature are based on commons-based peer production (and sharing) of energy, we argue that there is clear potential and capacity to build a platform for P2P energy sharing similar to the popular file sharing system BitTorrent (with requests, multi-source resource provision, opt-in, opt-out, "servers" akin to a BitTorrent tracker and more).

BitTorrent is a communication protocol that enables P2P sharing of data over the Internet. Instead of downloading files from only one specific server, files are distributed in a "swarm of hosts" within the BitTorrent system. Large files are split into small segments that are downloaded by peers requesting that file. Once the segment becomes available at a specific peer, the same peer becomes a source of data. One special element in this protocol is the "tracker", which is a special server that supports the P2P distribution by tracking the source of data segments and helping coordinate the transmissions as well as reconstruct the files. A more decentralized trackerless solution has also been proposed by using the "distributed hash table" method. BitTorrent uses the "tit-for-tat" rule to support cooperative behavior amongst its users.

PEM follows a similar premise: energy consumption becomes quantized into chunks of certain watt-hours with a predetermined duration (length) of certain minutes [43]. Therefore, multiplexed energy demand becomes possible, opening the possibility of managing energy loads in a decentralized manner based on service requests. However, unlike data exchanges, the physical electric grid imposes strict constraints to ensure a minimum level of power quality. Additionally, the way the electricity system is governed in each country adds another level of complexity.

Under the current regulations, PEM could be implemented virtually via SDENs, where the generation and storage can be aggregated via a set of new technologies (i.e., Virtual Power Plants and Virtual Batteries). At the demand side, PEM generates daily load profiles based on how flexible the energy consumption is. This becomes an inventory management problem with diverse classes of loads (e.g., thermal, transportation, cooking, dishwashers) as well as different and individualized profiles and preferences. PEM allows "energy packet requests" to coordinate them via packet multiplexing. Although there are many technical challenges involved in deploying this solution (like cyber-security, privacy and dependence on communication networks), similar technologies are commercially used in various countries, including households and industries[1].

SDENs and PEM align with the existing liberalized market with ancillary and balancing services. However, they also open up the possibility for democratizing electricity if governed as a commons. Within the dominant political economy, democratizing electricity would reduce profit maximization as more and more local actors enter the market [44]. So the incentives for new investments by the private sector would decrease hindering the proliferation of renewables via the electricity markets. Hence, a market mechanism for this type of energy production comes with inherent incompatibilities. While efforts to address the issues are underway, removing the market mechanism altogether may prove a more efficient solution.

In the capitalist framework, end users bear the cost for the long-distance transmission of electricity to their homes. For example, in Finland, the consumers are charged with electricity consumption and the network costs, which can be twice as expensive as the electricity consumption. If electricity is to be fully democratized, the network costs should be minimized including hardware, maintenance, and operational costs related to the long-distance transmission of electricity to the end user. But, how to minimize these costs?

Within liberalized markets, the demand management in minute timescale cannot be performed by the distribution network operator (DNO). Worse still, the retailers are many and can be located elsewhere than in the region of the DNO. The retailers with their associated ancillary service providers are the ones "controlling demand", and the DNOs and the transmission system operator (TSO) need to operationally handle this. Therefore, in our view, the (political) decision to minimize network costs is to (re)couple the demand management in all scales under the DNO, and thereby remove the need for (for-profit) retailers. Further, regulations should enable microgrids, i.e., localized energy grids that can function independently from the centralized grid. We envision a full-scale commons-based energy system utilizing networked P2P microgrids whose management is virtualized and based on PEM targeting self-sufficiency and electricity sharing (Figure 1). In this way, we have our first vision of an Energy Internet-based commons paradigm for the delivery of electricity.

[1] For example, see packetizedenergy.com and vps.energy.

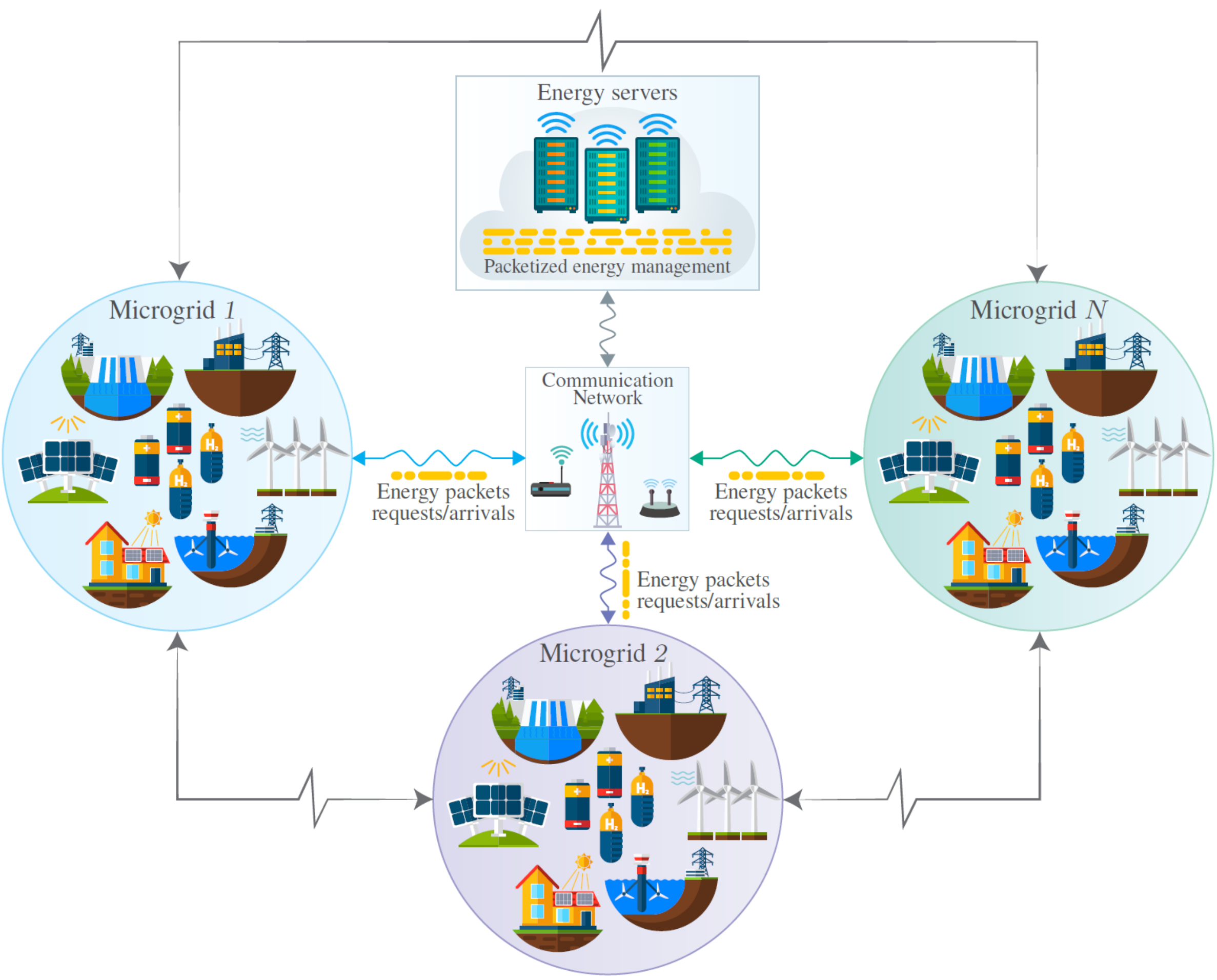

*Figure 1. A commons-oriented Energy Internet of networked microgrids. In this software defined energy network (SDEN), multiple microgrids (small local, often independent, grids) interact with each other to share electricity as a commons. Their interactions are managed and optimized using packetized energy management (PEM) through a communications network infrastructure designed on similar principles as the Internet.*

### *4.3 The energy commons pathway*

Without going into too much detail on demand-side management based on price, we suggest that SDEN and PEM enable the management of the collective supply, storage and demand of a group of consumers in minute time-scale. Efficient management is important because although there is enough energy to be harnessed, it is necessary to collectively coordinate the consumption so that availability does not become scarce for some persons, or at certain times of the day.

Our technical vision of an Energy Internet of networked microgrids that is oriented towards the commons is the following. A hierarchical management based on Home Energy Management Systems (HEMSs) would handle the flexible loads at the household level. These HEMSs aggregate the requests from the different flexible appliances and classify them based on priority levels (e.g., a laundry machine may have priority over charging an e-bike battery) and type of demand (e.g., interruptible like space heating devices, or uninterruptible like dishwashers). The HEMSs would be designed to work within the microgrids they are connected to, in a commons-based approach, so that renewable generation, flexible loads and storage capabilities are virtually aggregated following PEM.

Individual preferences and historical forecasts are considered to schedule the individual energy consumption based on requests and manage the storage elements. The goal of the optimization function carried out by an “energy server” (the entity that manages the microgrid) is to maximize self-sufficiency among the group associated with a specific microgrid community. In other words, the goal is to keep the microgrid as independent as possible, focusing on energy needs rather than “price signals" or “profit”. However, microgrids may also be interconnected so that they can interchange electricity when needed, creating a “hyper-energy server,” whose goal is to handle the microgrid requests. In this case, the energy server could be viewed as the equivalent of today’s DNO and the hyper-energy server as the equivalent of today’s TSO.

An important aspect of the proposed approach to the Energy Internet is the P2P sharing of energy resources by all the community members. The end users in a microgrid not only would produce energy locally using renewable resources, they would also share the produced energy with their neighbours. Both production resources and costs would be shared by the community to achieve overall social benefit. A self-contained and self-sufficient local energy supply system characterized by distributed renewable energy generation would be responsible for supplying energy to the community. The produced energy and system resources, costs, and benefits would be distributed among all the stakeholders, such as end users and governing/other entities, based on mutually agreed rules. Such a community-based renewable energy sharing is often called a community microgrid.

Community microgrids have been explored in the literature both from a social sciences perspective [45, 46, 47] and technical perspectives [48]. Numerous projects have also been implemented globally, but on a rather limited scale, curbed by capitalist regulatory and economic barriers [49, 45]. Moreover, many technical solutions focus on economic optimization by either working within the current electricity market framework or by imitating its structure to create localized markets [50]. Thus, the general approach is to not disrupt current market structures as much as possible.

Since these market structures are based on competition, co-operative approaches are often relegated to regions that either do not have any electricity access or do not have liberalized open competitive markets [51]. We argue that co-operative approaches, which focus on increased social welfare within the framework and values of a commons-oriented Energy Internet, are important disruptors and key enablers of renewable energy proliferation. Such radical approaches are arguably required if we are to move faster toward a cleaner and sustainable planet.

The technical challenges in this scenario are similar to the ones faced today. Namely less inertia in the system because of generation from non-rotating generators; need for adaptive protection schemes to allow variability of "power flows" due to renewable generation at distribution level; and coordination of storage facilities. However, a commons-oriented Energy Internet would offer a less complex operation by decreasing the layers of complexity from the current market structure, and by removing the "temptation" for profit-making speculation and consequently the constructed scarcity of an almost abundant good.

The idea, the current status, and the challenges of the Energy Internet are thoroughly reviewed by one of the authors and his colleagues [42]. In effect, current research has focused on building efficient Energy Internet-based solutions for realizing an efficient energy system within a microgrid. The economic aspects of implementing community microgrids and interconnected microgrids, i.e., how the members of a community can cooperate and share the resulting benefits fairly among themselves, have been discussed by another author of this paper [52]. It remains challenging to practically implement the Energy Internet solutions because of hurdles often stemming from regulatory barriers in capitalist market structures, non-participation from network operators, or community acceptance itself [52].

These hurdles mean that the research on developing these solutions into a commons-oriented paradigm is still at a nascent stage. This is compounded by the fact that the co-operative P2P sharing aspect threatens existing market structures. Unfortunately, most of the current focus is on profit maximization rather than on promotion of social welfare [52]. Nevertheless, researchers, including us, are moving toward developing Energy Internet-based solutions to enable a holistic sharing of energy resources within and between multiple microgrids on the basis of social welfare concepts such as maximizing renewable energy proliferation, reducing $CO_2$ emissions footprint, and achieving self-sustainability [53].

The aim of these new approaches should be to bypass the existing market structure. The new structure technically would allow for a balance between the individual needs of a shared good and orchestrating/scheduling the energy consumption as a commons with a guaranteed quality of service. We would neither redesign the wheel nor discard technical developments in electricity networks. Instead, we propose a feasible solution to rethink the energy systems in an era dominated by low-marginal cost electricity. This would eliminate the political and economic contingencies that complicate or jeopardize the system operations, often leading to unfair access in the energy supply.

## 5. Conclusions: Challenges and future horizons

We argued that most of the proposed solutions to the energy conundrum are hindered by a business-as-usual approach from a political economy perspective. We thus introduced a commons-oriented Energy Internet that may be a radical sustainable alternative to energy production and consumption. A commons-oriented Energy Internet is technically feasible given today's technological level. However, it requires a transition towards a new political economy framework centered around the commons. After all, the technical domain reflects the background social elements of the system:

individual users share their resources so that all can be supplied when the individual need appears. Likewise, individual microgrids share their resources so that all microgrids can have energy available when needed. No one actually "owns" the generated output since energy in this technical system is governed by all as a commons. As we argued, the technology reflects the mode of production from which it emerges.

So, the current social and economic arrangements would need to be reconceptualized to enable such a technological trajectory both politically and practically. Significant investments would need to take place for the infrastructure that would allow mass individual small-scale production to be possible. Maintaining that infrastructure would also be required. Although subsidies in the energy sector is a highly complex topic, with some providing support for marginalised social groups, an important percentage of those are directed towards sustaining structures that are entirely harmful both socially and environmentally [54]. The resources invested in such subsidies could be redirected into enabling civilians to produce energy as well the creation of structures to support the model like local and national governance associations, while providing support for the social groups that are adversely affected by the shift. The state, which supports current energy models with subsidies, could fund the infrastructure costs via energy commons as "green energy" programs, in a similar way to how taxation seeks to distribute public goods equally. Our proposal to shift such subsidies towards those costs may be a basic preliminary proposal yet an arguably good place to start researching towards this line of inquiry. We believe that a pathway towards exemplifying this direction could initially see the socialized costs of infrastructure distributed among the members of the community in various ways. Moreover, to evaluate the potential long-term social and environmental impacts and benefits of the proposed solution, it is important to use quantitative and qualitative assessment tools such as the life-cycle-assessment, the matrix of convivial technology or the open-o-meter [55].

We thus have presented an alternative mode of production as well as a potential system for energy production that may be better suited to tackle the current issues society faces. For the feasibility of such alternatives, bold and radically rethink is needed on how to produce and consume energy. And we also need to reconceptualize the modern way of living, hence the importance of coupling energy proposals with wider systemic change away from a paradigm of constant growth. At this point in time, given the severity of our situation, these solutions should not be considered utopian dreams but realistic, if not necessary objectives.